\documentclass[12pt]{article}
\usepackage{}
\usepackage{epsfig}

\usepackage{a4}
\usepackage{amsmath}

\usepackage{graphicx}  
\usepackage{epsfig}
\usepackage{amsmath}
\usepackage{amsfonts} 
\usepackage{amssymb}
\usepackage{ifthen}

\begin{document}
\title{On the topological charge of $SO(2)$ gauged Skyrmions in $2+1$ and $3+1$ dimensions}
\author{{\large Francisco Navarro-L\'erida,}$^{1}$
{\large Eugen Radu}$^{2,3}$
and {\large D. H. Tchrakian}$^{3,4}$ 
\\ 
\\
$^{1}${\small Dept.de F\'isica Te\'orica, Ciencias F\'isicas,}\\
{\small Universidad Complutense de Madrid, E-28040 Madrid, Spain}\\  
$^{2}$ 
{\small Center for Research and Development in Mathematics and Applications,}
\\
 {\small 
 (CIDMA) Campus de Santiago, 3810-183 Aveiro, Portugal
}\\ 
$^{3}${\small School of Theoretical Physics, Dublin Institute for Advanced Studies,}
\\{\small 10 Burlington Road, Dublin 4, Ireland }
\\
  $^{4}${\small  Department of Computer Science, NUI Maynooth, Maynooth, Ireland}}

\date{\today}
\newcommand{\dd}{\mbox{d}}
\newcommand{\tr}{\mbox{tr}}
\newcommand{\la}{\lambda}
\newcommand{\om}{\omega}
\newcommand{\ka}{\kappa}
\newcommand{\ta}{\theta}
\newcommand{\f}{\phi}
\newcommand{\vf}{\varphi}
\newcommand{\vr}{\varrho}
\newcommand{\F}{\Phi}
\newcommand{\al}{\alpha}
\newcommand{\bt}{\beta}
\newcommand{\ga}{\gamma}
\newcommand{\de}{\delta}
\newcommand{\si}{\sigma}
\newcommand{\Si}{\Sigma}
\newcommand{\bomega}{\mbox{\boldmath $\omega$}}
\newcommand{\bsi}{\mbox{\boldmath $\sigma$}}
\newcommand{\bchi}{\mbox{\boldmath $\chi$}}
\newcommand{\bal}{\mbox{\boldmath $\alpha$}}
\newcommand{\bpsi}{\mbox{\boldmath $\psi$}}
\newcommand{\brho}{\mbox{\boldmath $\varrho$}}
\newcommand{\beps}{\mbox{\boldmath $\varepsilon$}}
\newcommand{\bxi}{\mbox{\boldmath $\xi$}}
\newcommand{\bbeta}{\mbox{\boldmath $\beta$}}
\newcommand{\ee}{\end{equation}}
\newcommand{\eea}{\end{eqnarray}}
\newcommand{\be}{\begin{equation}}
\newcommand{\bea}{\begin{eqnarray}}

\newcommand{\ii}{\mbox{i}}
\newcommand{\e}{\mbox{e}}
\newcommand{\pa}{\partial}
\newcommand{\Om}{\Omega}
\newcommand{\vep}{\varepsilon}
\newcommand{\bfph}{{\bf \phi}}
\def\theequation{\arabic{equation}}
\renewcommand{\thefootnote}{\fnsymbol{footnote}}
\newcommand{\re}[1]{(\ref{#1})}

\newcommand{\eins}{1\hspace{-0.56ex}{\rm I}}
\newcommand{\R}{\mathbb R}
\newcommand{\C}{\mathbb C}
\newcommand{\p}{\mathbb P}
\renewcommand{\thefootnote}{\arabic{footnote}}

\maketitle


\bigskip

\begin{abstract}
The question of the dependence of the topological charge $q$ of a gauged Skyrmion, on the gauge field, is studied quantitatively. 
Two examples, both gauged with $SO(2)$ are studied and contrasted:
{\bf i)} The $O(3)$ model in $2+1$ dimensions, and {\bf ii)} The $O(4)$ model in $3+1$ dimensions.
In case {\bf i)}, where the (usual) Chern-Simons (CS) term is present, the value of
$q$ changes sign, going through zero. This evolution is tracked by a parameter characterising the solutions in the given theory.
In case {\bf ii)}, in which dimensions no CS density is available, the evolution of $q$ is not observed.

\end{abstract}
\medskip
\medskip

\section{Introduction}
The topological charge $q$ of gauged Skyrmions presents peculiarities that are absent in gauged Higgs solitons.
While in the latter case this quantity always equals the winding number of the Higgs field, in the case of
Skyrmions the value of $q$ in general may depart from the winding number, or the ``baryon number''.

The topological charge for the $SO(2)$ gauged $O(4)$ (Skyrme) model in $3+1$ dimensions is defined in Ref.~\cite{Callan:1983nx}
and in \cite{Piette:1997ny,Radu:2005jp}, and, for the $SO(2)$ gauged $O(3)$ (planar Skyrme) model in $2+1$ dimensions
in Ref.~\cite{Schroers:1995he}. The topological charges for the $SO(D)$ gauged $O(D+1)$ sigma model on $\R^D$ ($D=2,3,4$) were defined in
\cite{Tchrakian:1997sj}, and the soliton of the  $SO(3)$ gauged $O(4)$ sigma model on $\R^3$ was first reported in \cite{Arthur:1996np}.
More detailed studies of the latter solitons were given in \cite{Brihaye:1998vr}, while solitons of the  $SO(4)$ gauged $O(5)$ sigma model on $\R^4$
were reported in \cite{Brihaye}.
The most comprehensive and recent definitions of the generic topological charges of gauged Skyrmions are given 
in Appendix {\bf B} of Ref.~\cite{Tchrakian:2015pka}.

With the exception of the work in Ref.~\cite{Callan:1983nx}, where attention was focused on the decay of the baryon number~\footnote{
Well known works in this direction are, $e.g.$, \cite{DHoker:1984wbw,Rubakov:1984ba}, employ instead $SU(2)$ gauging that does not
result in topologically stable solitons.}, the works of \cite{Piette:1997ny,Radu:2005jp,Schroers:1995he,Tchrakian:1997sj,Arthur:1996np,Brihaye} are
concerned with the construction of topologically stable solitons.
Here, we are concerned with the $SO(2)$ gauged $O(3)$ and
$O(4)$ sigma models in $2+1$ and $3+1$ dimensions, respectively. But unlike in the works of Refs.~\cite{Schroers:1995he,Piette:1997ny,Radu:2005jp},
our focus here will be the evolution of the topological charge, $e.g.$, its possible decay.\footnote{
$SO(2)$ gauged Skyrmions in $2+1$ and dimensions have been studied extensively
(see Refs.~\cite{Schroers:1995he,Navarro-Lerida:2016omj,Ghosh:1995ze,Kimm:1995mi,Arthur:1996uu}
 and more recently in \cite{Samoilenka:2016wys,Samoilenka:2015bsf}). The
$SO(2)$ gauged Skyrmions in $3+1$ dimensions have been quantitatively studied mainly in
Refs.~\cite{Piette:1997ny,Radu:2005jp} (see also the recent analytical results in Ref. \cite{Canfora:2018clt}).
}

\medskip

In odd dimensional spacetimes, where the Chern-Simons (CS) term is defined, one finds that the effect of this CS
dynamics results in the mass/energy of the static solitons both to increase and to decrease with increasing
global charge (electric charge and spin). This was clearly demonstrated in Section {\bf 4}
of Ref.~\cite{Navarro-Lerida:2016omj} in $2+1$ dimensions, where the effect of Chern-Simons dynamics were studied also in
other $2+1$ dimensional models. This evolution of the mass/energy of the static solitons was tracked by a parameter characterising the solution in the $given$ theory,
which is strictly due to the CS dynamics. In the present work we show that this same dynamical effect results in the evolution of the value of the
topological charge $q$, changing sign and passing through zero, in the given theory. Such a solution cannot be found in the absence of the CS term~\cite{Navarro-Lerida:2016omj}.
This is the main result of the first part of the present work, namely the study of the $SO(2)$ gauged $O(3)$ model,
with emphasis on the issue of topological charge $q$. 
(A detailed study of generic $SO(2)$ gauged Skyrmions is given in \cite{Francisco}.)

In even dimensional spacetimes however, no CS density is defined. There is nonetheless a class of Skyrme-CS (SCS)
densities satisfying the properties of the usual CS density in all dimensions, including even dimensions. These are
introduced in Ref.~\cite{Tchrakian:2015pka}. 
It happens that in the present example  this quantity  
vanishes under the imposition of appropriate symmetries, 
and in the absence of any Chern-Simons type dynamics we are limited to
studying the model considered in Refs.~\cite{Piette:1997ny,Radu:2005jp}. 
There, it was found that the mass/energy
of the soliton of the gauged system increases monotonically with increasing electric (global) charge. 
Since no parameter characterising the solutions is available,
the coupling strength of the Maxwell field, $\la_0$ in \re{Lso2} below, was varied in \cite{Piette:1997ny}.
This clearly changes the theory being considered, but since in any given theory the value of both the energy and the electric charge $Q_e$ depend on $\la_0$,
this enabled~\cite{Piette:1997ny} the tracking of the energy with increasing $Q_e$.
Here by contrast, we consider the dependence of the topological charge on $\la_0$ 
and show that  $q$ remains constant.

\medskip
 
This paper is organised as follows.
In Section {\bf 2} below we present our quantitative results on the $SO(2)$ gauged Skyrmion in $2+1$ dimensions, and in 
Section {\bf 3} our results on the $SO(2)$ gauged Skyrmion in $3+1$ dimensions. In Section {\bf 3} we summarise our results and point out to further developments.

\section{$SO(2)$ gauged Skyrmions in $2+1$ dimensions}
Ever since Schroers' construction of solitons~\cite{Schroers:1995he} of the $U(1)$ gauged planar Skyrme model, there has been considerable
interest~\footnote{See Ref.~\cite{Navarro-Lerida:2016omj} and citations there.} in this area. The dynamics of the Abelian field in Ref.~\cite{Schroers:1995he}
was described by the Maxwell density, subsequently the latter was replaced by the Chern-Simons (CS) density~\cite{Ghosh:1995ze,Kimm:1995mi,Arthur:1996uu}. More
recently, the effect of Chern-Simons dynamics on the combined Maxwell--CS $O(3)$ sigma model was studied in Ref.~\cite{Navarro-Lerida:2016omj}. What was found there
(in Section {\bf 4}) is that the mass/energy of the soliton decreased with increasing global charges -- electric charge and angular momentum -- in some ranges
of the parameters. This effect is absent, in the absence of the CS term in the Lagrangian, and is strictly a result of the Chern-Simons dynamics.

In the present note, we pursue further consequences of the presence of the Chern-Simons term. 
The new effect in question is the evolution of the topological charge
due to the gauging of the the $O(3)$ sigma model. While this is not strictly a dynamical effect, it is again predicated on the presence of a CS term in the
Lagrangian. The precise mechanism of this is the role that the CS term plays 
in establishing the topological lower bound (Belavin inequalities). 
This, and
other details will be exposed in a future extended work.

The Skyrme model on $\R^2$ is described by the scalar $\f^a=(\f^\al,\f^3)\ ,\quad |\f^a|^2=1\ ,$ and the $SO(2)$
gauging prescription is~\footnote{The prescription for gauging the $O(D+1)$
Skyrme scalar $\f^a\ ,\ a=1,2,\dots,D+1$, on $\R^D$ with 
 gauge group 
$SO(N)$, $2\le N\le D$, is summarised most recently in Appendix {\bf B} of Ref.~\cite{Tchrakian:2015pka}.}
\bea
D_\mu\f^{\al}&=&\pa_\mu\f^{\al}+A_\mu(\vep\f)^{\al}\ ,\quad(\vep\f)^{\al}=\vep^{\al\bt}\f^\bt, 
\label{cal}
\\
D_\mu\f^3&=&\pa_\mu\f^3\label{c3}\ ,\quad\mu=i,0\ ,\ i=1,2\,,
\eea
$A_\mu$ being the $SO(2)$ connection.

The topological charge density for the $SO(2)$ gauged Skyrme model is 
\cite{Tchrakian:1997sj},
\cite{Tchrakian:2015pka},
\cite{Francisco}
\be
\label{tc}
\varrho = \varrho_0+2\vep_{ij} \pa_{i} [(\phi^3-\upsilon)A_{j}] , 
\ee
independently of whether the dynamics is Maxwell or CS. In \re{tc}, $\vr_0$ is the winding number density whose volume integral is the ``baryon number'', $A_i\ ,\, i=1,2\ ,$ are the magnetic components of 
the Abelian connection, and $v$ is a real constant.

What is crucial here is that in the Maxwell gauged case the Belavin inequalities, which must be satisfied by the requirement of topological stability, force the constant $v$
to be equal to one, $v=1$, and as a consequence of the boundary condition $\f^3(\infty)=1$, the contribution of the Abelian field to the integral of $\vr$ disappears. What remains is the
integral of $\vr_0$, namely, the ``baryon number''. The situation is starkly different when the dynamics of the Abelian field is described by the Chern-Simons
term. In that case, it follows from the Bogomol'nyi analysis, that the constant $v$ in \re{tc} can take any value, including $v\neq 1$, as a consequence of which the topological charge will depart from the ``baryon number'',
quantitatively. It is our aim here to track the evolution of the topological charge, due to this mechanism.

To illustrate this mechanism briefly, consider the choice of potential $V \equiv V[\f^a]$ in the Lagrangian.
 In the case of Maxwell dynamics, this potential is fixed by the constraints
of the Belavin inequalities to be
\be
\label{potM}
V_{\rm M}=\frac12\la(1-\f^3)^2\,,
\ee
consistently with the boundary value $\f^3(\infty)=1$, when $V_{\rm M}[\f^3(\infty)]=0$.

The corresponding Bogomol'nyi analysis
in the case of Chern-Simons dynamics, 
which was presented in \cite{Arthur:1996uu} and in Appendix C 
of \cite{Francisco}, leads to the potential
\be
\label{potCS}
V_{\rm CS}=\frac{1}{32}\,\la\,\left(\frac{\eta}{\ka}\right)^2|\f^\al|^2(\f^3-\upsilon)^2\ ,\quad\la>0\,.
\ee
(In \re{potM}-\re{potCS}, $\la\ ,\eta$ and $\ka$ are real constants.) The striking property of the potential \re{potCS} is that the (finite energy) condition $V_{\rm CS}[\f^a(\infty)]=0$
can be satisfied for any value of the real constant $v$, since at spatial infinity
\be
\label{vneq0}
|\f^3|^2\to 1\quad\Rightarrow\quad|\f^\al|^2\to 0\ ;\quad\al=1,2\,.
\ee
This is precisely the situation where the topological charge will depart from the winding number $n$, namely from the ``baryon number''. In practice, $v=0$ is the most convenient choice.

The Chern-Simons--Skyrme (CS-S) model studied in Ref.~\cite{Arthur:1996uu},
can be augmented by the Maxwell term without invalidating the topological inequalities, to yield the Maxwell--CS--S model described by the Lagrangian
\be
\label{MCSSLag}
{\cal L}=-\frac14 F_{\mu\nu}^2+\ka\vep^{\la\mu\nu}A_\la F_{\mu\nu}
-\frac18\tau D_{[\mu}\f^aD_{\nu]}\f^b+\frac12\eta^2|D_\mu\f^a|^2-\eta^4\,V_{\rm CS}[\f^a],
\ee
considered here.

The Lagrangian \re{MCSSLag} is not the minimal one that follows from the Belavin inequalities. It has been further augmented by the quartic kinetic Skyrme term with coupling $\tau$.
The reason for this is our requirement that after gauge decoupling, the system support Skyrmions with winding number $n$.

To analyse the system \re{MCSSLag} numerically, we subject it to azimuthal symmetry
\bea
\label{axO3}
\f^{\al}&=&\sin f(r)\,n^{\al}\ ,\quad \f^3=\cos f(r)\ ,\quad
n^{\al}=\left(
\begin{array}{c}
\cos n\ta\\
\sin n\ta
\end{array}
\right),
\\
A_i&=&\left(\frac{a(r)-n}{r}\right)(\vep\hat x)_i\ ,\quad A_0=b(r)\,,\label{Maxax}
\eea
where $\ta$ is the azimuthal angle and $n$ is the winding number of the Skyrme scalar.

\begin{figure}[h!]
\centering
\includegraphics[height=3in]{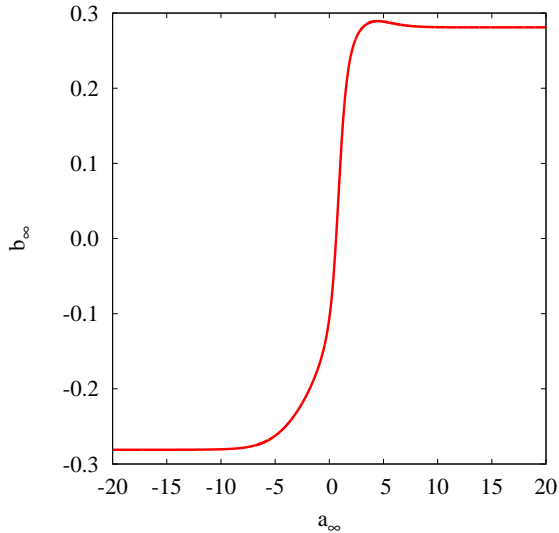}  
\caption{
$b_\infty$ vs. $a_\infty$ for vortices with $n=1$, $\lambda=1.6$, $\eta=1$, $\kappa=1$, and $\tau=1$. 
} 
\label{fig_binf_vs_ainf}
\end{figure} 
 
We have analysed the reduced the one dimensional Lagrangian descended from \re{MCSSLag} subject to 
\re{axO3},
\re{Maxax}, numerically.
Also, subject to
the above Ansatz,
the volume integral of $\vr$ (as given by \re{tc}), namely the topological charge, can be readily evaluated as\footnote{
Subject to the Ansatz \re{axO3}-\re{Maxax},  
the electric charge $Q_e$ is 
$
 Q_e=8\pi\kappa (n - a_\infty),  
$
while
 the total mass/energy $E$ and angular momentum $J$
 are computed as integrals of the $T_{00}$ and $T_{i0}$
components of the energy momentum tensor \cite{Navarro-Lerida:2016omj}.
}
\be
\label{fintc}
q=-\frac{1}{8\pi}\int\vr_0\,d^2x=\frac12(n+a_\infty)\, ,
\ee
where $a_\infty=a(\infty)$.

\begin{figure}[h!]
\begin{center}
\includegraphics[height=3in]{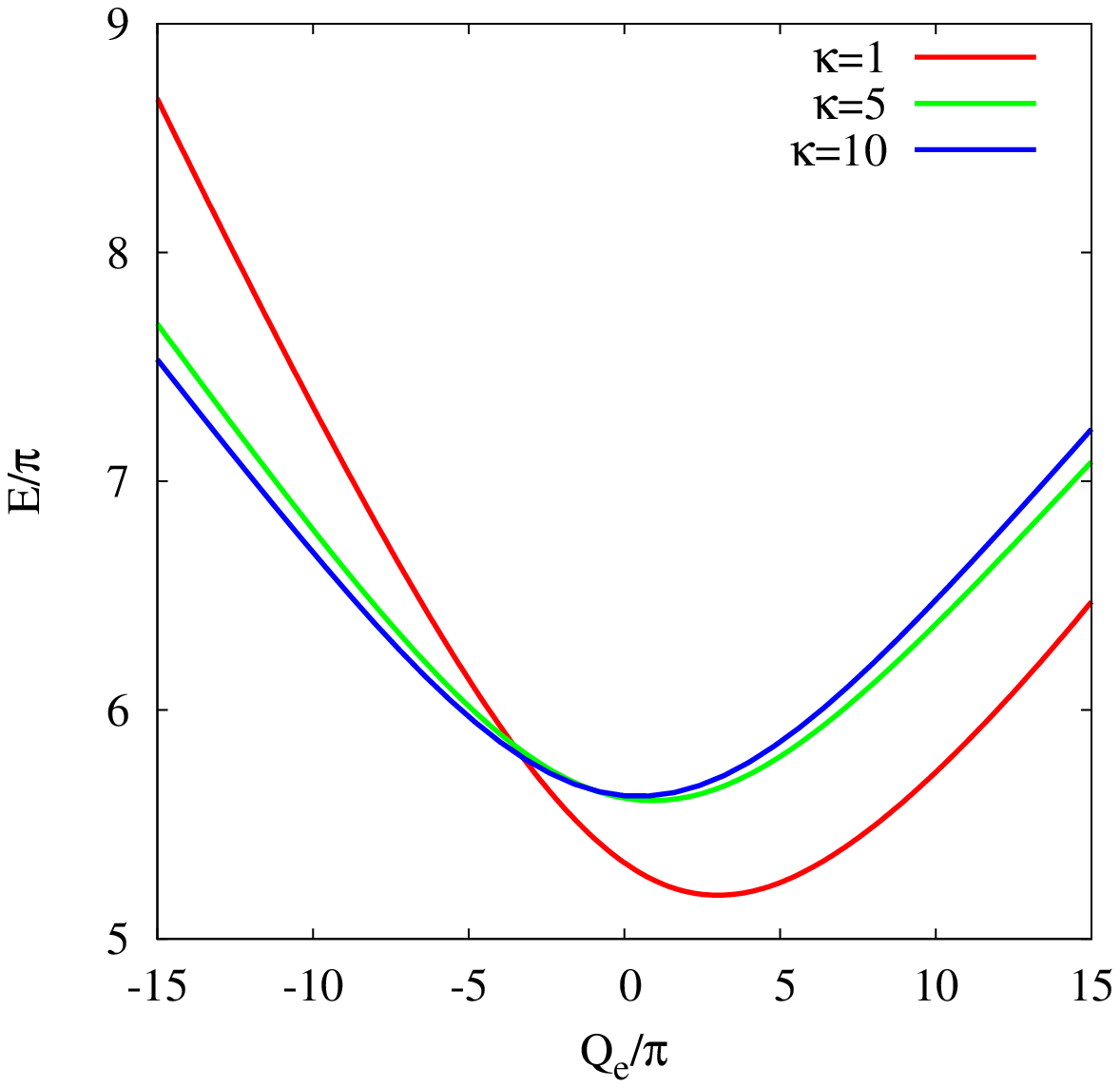}
\includegraphics[height=3in]{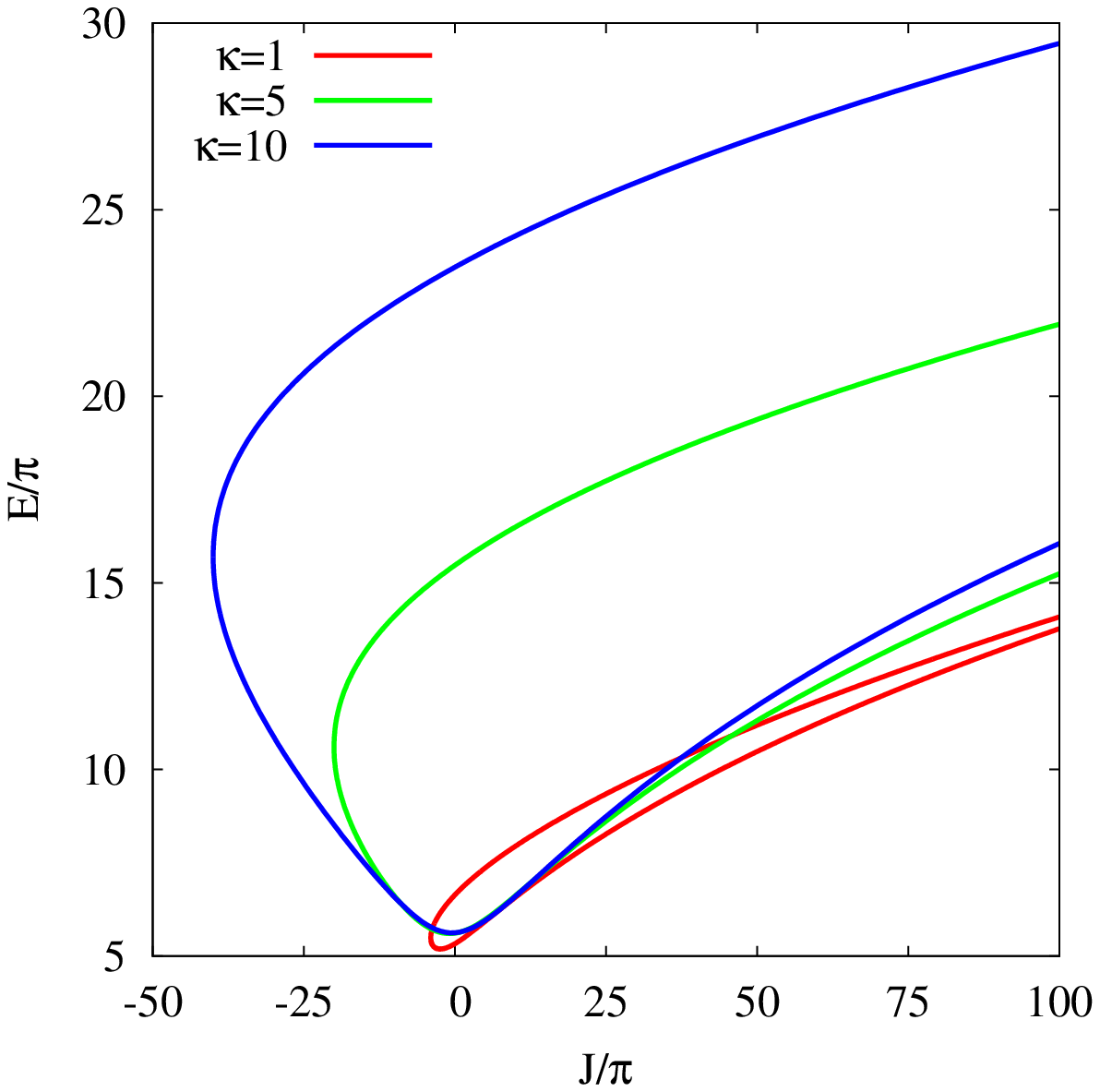}    
\caption{ {\it Left panel:} Energy $E$ $vs.$ electric charge $Q_e$ for vortices with $n=1$, $\lambda=1.6$, $\eta=1$, $\tau=1$, and several values of $\ka$. {\it Right panel:} Energy $E$ $vs.$ angular momentum $J$  for vortices with $n=1$, $\lambda=1.6$, $\eta=1$, $\tau=1$, and several values of $\ka$.} 
\label{figs2}
\end{center}
\end{figure} 
\begin{figure}[h!]
\centering
\includegraphics[height=3in]{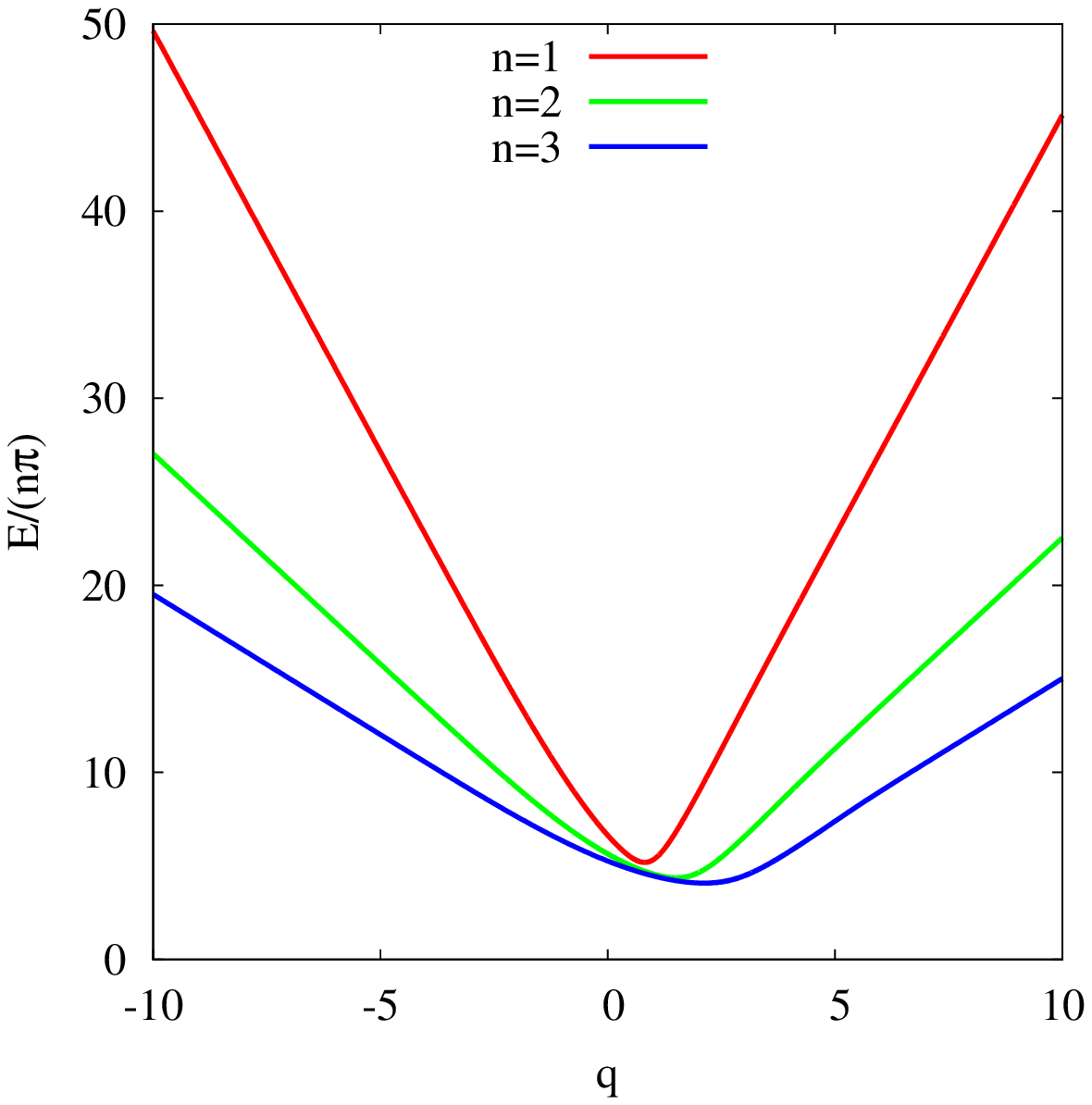}
\includegraphics[height=3in]{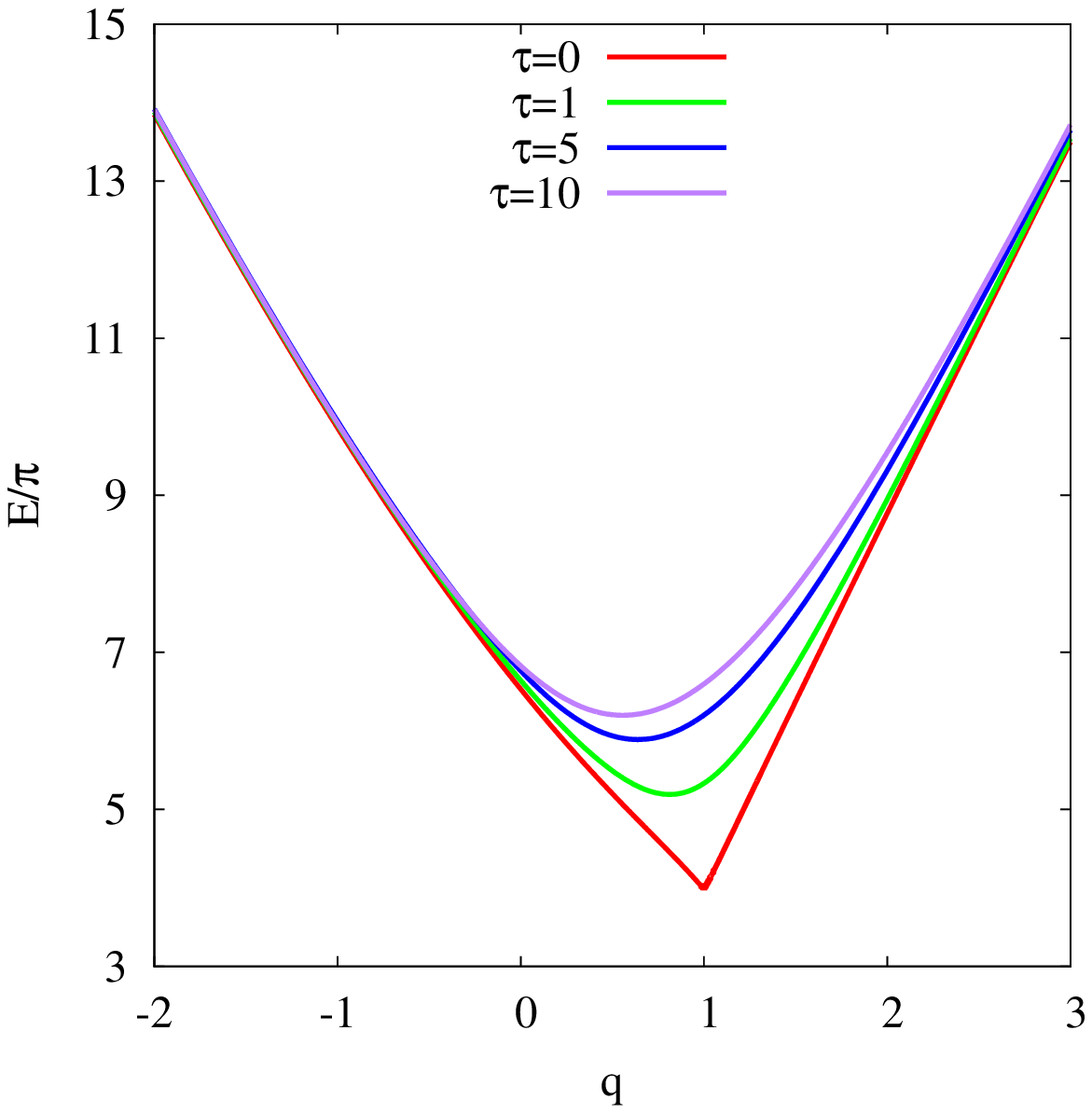}    
\caption{ {\it Left panel:}
Energy  $E$ $vs.$
 topological charge $q$ for vortices with $n=1,2,3$, $\lambda=1.6$, $\eta=1$, $\kappa=1$, and $\tau=1$.
{\it Right panel:}
Same for vortices with $n=1$, $\lambda=1.6$, $\eta=1$, $\kappa=1$, and  several values of $\tau$.} 
\label{fig_E_vs_q_sev_n}
\end{figure} 

We see from the above relation that solutions characterised by $a_\infty<0$ result in the dissipation of the topological charge. Thus, constructing solutions with $a_\infty<0$,
one can track the evolution of the topological charge, resulting also in the annihilation of $q$. The necessary handle on $a_\infty$ needed for this results precisely
from the same mechanism encountered in Ref.~\cite{Navarro-Lerida:2016omj}, where the (magnetic) quantity $a_\infty$ changes with the (electric) $b_{\infty} (=b(\infty))$. This is
strictly a Chern-Simons (CS) effect, as the CS density features both the magnetic and electric fields.

We have constructed such solutions with $\ka\neq 0$,  
the relation of the asymptotic value of the electric function, $b_\infty$ to the magnetic counterpart $a_\infty$ being displayed in Figure~\ref{fig_binf_vs_ainf}.
We know already from our results in Ref.~\cite{Navarro-Lerida:2016omj} that due to this effect, the dependence of the mass/energy $E$ on the global charges is non-standard, namely the dependence on the electric charge
$Q_e$ and the angular momentum $J$ displayed respectively (left and right panels) in Figure~\ref{figs2}.

Here, we analyse further the relation of the topological charge $q$ and the energy $E$. In Figure~\ref{fig_E_vs_q_sev_n}
(left panel)
 we represent the energy per unit winding number $n$ versus the topological charge for three values of $n$ 
for the same parameters as in Figure.~\ref{fig_binf_vs_ainf}. 
For small values of the coupling strength $\tau$ of the (quartic) Skyrme kinetic term, the minimum of the energy occurs at values of the topological charge around the winding number. However, when higher values of $\tau$ are 
considered, the minimum of the energy occurs 
at values clearly different from the winding number. This is shown in Figure.~\ref{fig_E_vs_q_sev_n}
(right panel), 
where the minimum of the curve for $n=1$, $\tau=10$ is located at $q \approx 0.556$. Concerning the stability of these solutions, one would be tempted to state that most stable solution would correspond to that with least energy,
which according to Figure.~\ref{fig_E_vs_q_sev_n} (right panel) does not possess integer topological charge, in general.

 From Figure 3. (right panel), we see that in the absence of the Skyrme term ($\tau=0$) the energetically favoured configurations are those with topological charge equal to the winding number. But when the Skyrme term is introduced,
with increasing values of the coupling strength $\tau$, it appears that the energetically favoured solutions have topological charge progressively smaller than the winding number, the latter being the topological charge of the Skyrmion in the gauge
decoupling limit.  

\section{$SO(2)$ gauged Skyrmions in $3+1$ dimensions}

The stationary solutions of this system were studied in Refs.~\cite{Piette:1997ny} and \cite{Radu:2005jp}. 
In \cite{Piette:1997ny}, it was seen that the mass of the electrically charged
Skyrmion was larger than that of the electrically neutral one, while in \cite{Radu:2005jp} 
it was shown that these Skyrmions spin. That the mass of the electrically charged Skyrmion
is always larger than the electrically neutral one was seen in \cite{Piette:1997ny} 
by observing that these energies both depend on the Maxwell coupling $\la_0$ (in \re{Lso2} below),
thus plotting both these energies $versus$ $\la_0$ illustrates this fact, as seen in Figure \ref{fig4} (left panel).

That the mass increases monotonically with increasing electric charge seen in \cite{Piette:1997ny} is not surprising, since we know from our work in Ref.~\cite{Navarro-Lerida:2016omj}
that to reverse this trend can be achieved only by the introduction of Chern-Simons (CS) dynamics, and, in $3+1$ dimensions no (usual) CS density is defined. While there is a class
of densities introduced in Ref.~\cite{Tchrakian:2015pka} which share the properties of Chern-Simons that can be defined in even dimensional spacetimes, the particular one that applies here,
defined in terms of the Maxwell field and the $O(6)$ Skyrme scalar, actually vanishes when subjected to axial symmetry. Hence, in the absence of any Chern-Simons dynamics, we are
back to the study in Refs.~\cite{Piette:1997ny,Radu:2005jp}.

What we have done here, additionally to tracking the increase of mass with increasing electric charge and angular momentum in Refs.~\cite{Piette:1997ny,Radu:2005jp}, is to study the
effect of the electromagnetic field on the topological charge~\footnote{Strictly speaking this is the deformation of the topological charge by the gauge field, that provides a
lower bound on the energy of the gauged system.}.

The studied  $SO(2)$ gauged Skyrmion system is described by the Lagrangian
\be
\label{Lso2}
{\cal L}=
-\frac14\,\lambda_0 |F_{\mu \nu}|^2 + \frac12\,\lambda_1 |D_{\mu} \phi^a|^2
-\frac14\,\lambda_2 |D_{[\mu} \phi^a D_{\nu ]}\phi^b|^2
+\lambda_3 V(\phi^a) ,
\ee
where $\f^a=(\f^\al,\f^A)\ ;\ \al=1,2\ ;\ A=3,4$, with $|\f^a|^2=1$, Skyrme scalars of the  $O(4)$ sigma model, with
$V=1-\phi^4$  the usual Skyrme potential.
$\lambda_i\ (i=0,1,2,3),$ in \re{Lso2} are real numbers parameterising the coupling constants of the model.
Also,
the gauging prescriptions for $\f^a$ is
\bea
D_{\mu} \phi^{\al}&=&\pa_{\mu} \phi^{\al}+A_{\mu}(\vep\f)^{\al}\ , \qquad D_{\mu} \phi^A=\pa_{\mu} \phi^A\ ,\quad \al=1,2\ ,
\quad A=3,4.
\label{covf1}
\eea 
 
In the ungauged case,
the topological charge density of the Skyrmions is
\begin{eqnarray}
\label{ro0}
\vr_0= \epsilon_{ijk} \epsilon_{abcd} \partial_i \phi^a \partial_j \phi^b \partial_k \phi^c \phi^d~.
\end{eqnarray} 
The natural generalization of
this expression
in the presence of gauge fields
is
found by replacing 
partial derivatives with gauge derivatives,
%
\begin{eqnarray}
\label{rog}
\vr_G= \epsilon_{ijk} \epsilon_{abcd} D_i \phi^a D_j \phi^b D_k \phi^c \phi^d ,
\end{eqnarray}
which,  
however,  does not leads to a conserved charge.

The conserved, gauge invariant generalization of  (\ref{ro0}) 
is the sum of two term\footnote{This definition is that used earlier in Refs.~\cite{Callan:1983nx} and \cite{Piette:1997ny}.}
%
\begin{eqnarray}
\label{def1}
\vr=\vr_0+\partial_i \Omega_i, ~~{\rm with}~~ 
\Omega_i=3 \epsilon_{ijk}\vep^{AB}A_j\partial_k \phi^A\phi^B,
\end{eqnarray} 
%
%
%
which can be 
can be written alternatively in a manifestly gauge invariant form
\begin{eqnarray}
\label{def2}
\vr=\vr_G+W,~~{\rm  with}~~
 W=\frac{3}{2}\epsilon_{ijk}F_{ij}(\vep^{AB}A_j \phi^BD_k\phi^A).
\end{eqnarray}
As such, the topological charge $q$ consists in the sum of two terms,  
encoding the contribution of $\vr_G$ and $W$ in (\ref{def2}) 
\begin{eqnarray}
\label{gdin}
q=-\frac{1}{2\pi^2}\int d^3 x \vr=q_G+q_W,
\end{eqnarray}
with $q_G$ the $bare$ charge (as found by integrating directly the density (\ref{rog})),
and $q_W$ a $compensating$ contribution (from (\ref{def2})).
Moreover, as seen from (\ref{def1}) 
the topological charge is still an integer, as long as there are no
singularities and the surface terms from $\Omega_i$ vanish.

Imposing axial symmetry on the Maxwell connection $A_\mu=(A_i,A_z.A_0)$, we have the Ansatz
\be
\label{abel}
A_{i}=-\left(\frac{a+n}{\rho}\right)\,(\vep\hat x)_{i}\ ,\quad A_z=0\ ,\quad A_0=-b,
\ee
where $a=a(\rho,z)\ ,\ b=b(\rho,z)$, and, $\rho^2=|x_i|^2$, $i=1,2$.

The solutions constructed are typified by the asymptotic values
\be
\label{asymval}
A_{i}=-\left(\frac{a_\infty+n}{\rho}\right)\,(\vep\hat x)_{i}\ ,\quad A_0=-b_\infty+\frac{Q}{\sqrt{\rho^2+z^2}}~,
\ee
where $Q$ is the electric charge and $b_\infty$ is a free constant characterising the given solution
(the electrostatic potential). Also, in contrast with the solutions in $2+1$
dimensions constructed in the previous Section, 
there is no freedom in the asymptotic value of the magnetic potential, with
$a_\infty=-n$ for all solutions. Again we are using the notation $a_\infty=a(\infty)$ and $b_\infty=b(\infty)$.

The corresponding Ansatz for the $O(4)$ scalar $\f^a$ is
\bea
\f^{\al}&=&R\,n^{\al}\ ,\quad\f^3=S\ ,\quad\quad\ \f^4=T\label{fa1},
\eea
where $R,S$ and $T$ are functions of $\rho$ and $z$,
subject to the constraint
$R^2+S^2+T^2=1$.
Also, $n^{\al}$ is the unit vector in the subplane $(x^1,x^2)$, with vorticity $n$.

Subject to symmetry \re{abel}-\re{fa1}, the topological charge density \re{def1} 
(or (\ref{def2}))
reduces to the two dimensional density
\bea
\label{RSTa}
\vr&=&\frac{1}{\rho}\bigg\{2nR\left(R\,\pa_{[\rho}S\,\pa_{z]}T+S\,\pa_{[\rho}T\,\pa_{z]}R+T\,\pa_{[\rho}R\,\pa_{z]}S\right)
\nonumber\\
\qquad\qquad\qquad &&-\left(S\,\pa_{[\rho}a\,\pa_{z]}T-T\,\pa_{[\rho}a\,\pa_{z]}S\right)-2(a+n)\pa_{[\rho}S\,\pa_{z]}T
\bigg\},
\eea
\label{norm}
%
%
 \begin{figure}[h!]
\begin{center}
\includegraphics[width=0.495\textwidth]{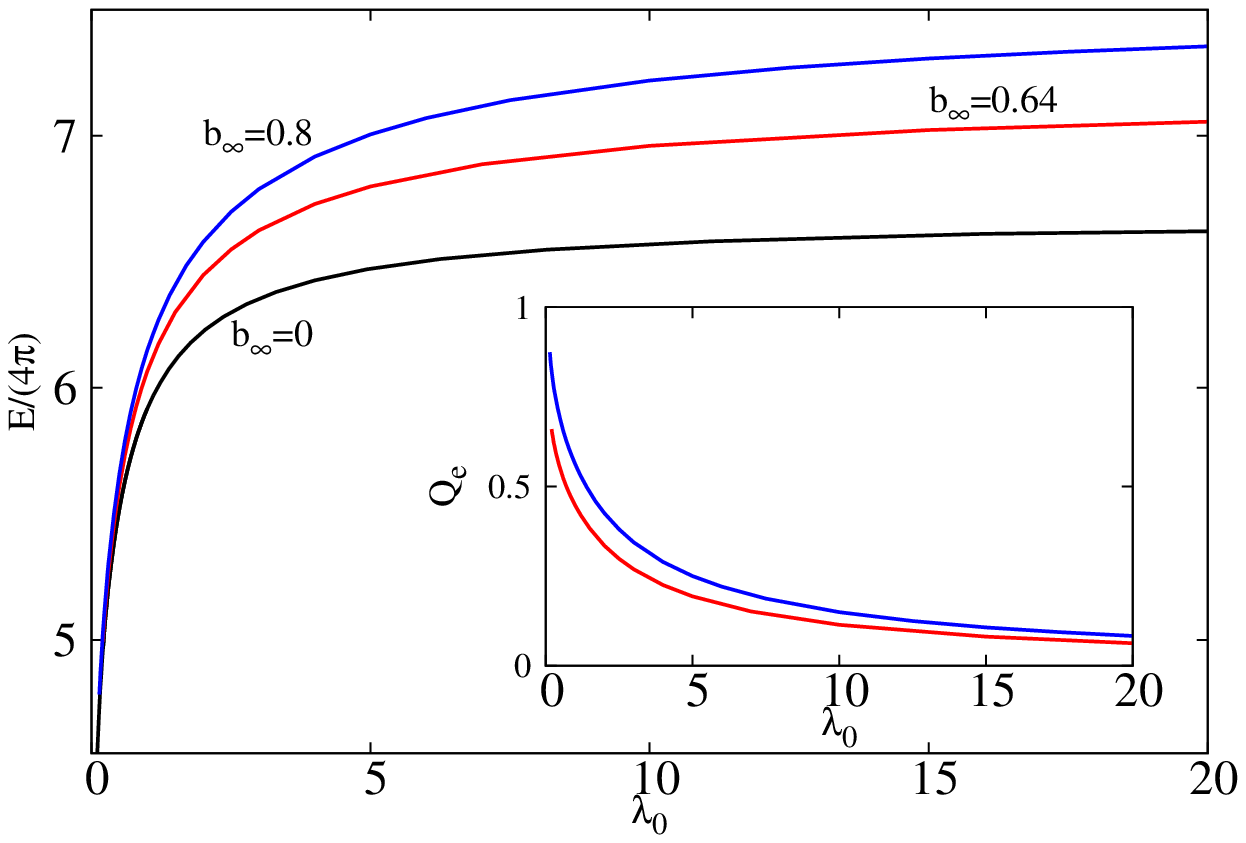} 
\includegraphics[width=0.495\textwidth]{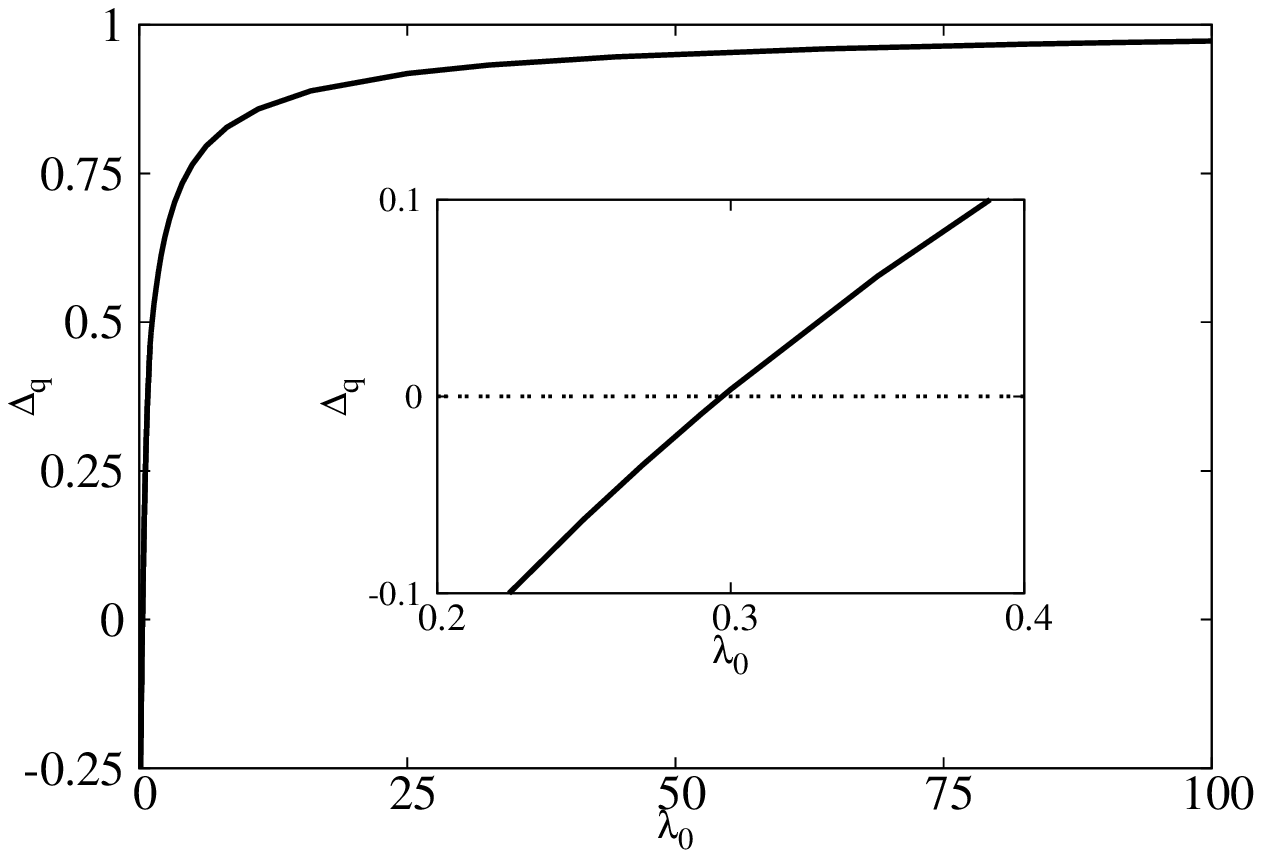}
\caption{\small{
{\it Left panel:} The energy $E$ and the electric charge $Q_e$ are shown 
$vs.$ 
the coupling constant 
$\lambda_0$ for SO(2) gauged Skyrmions in $3+1$ dimensions, 
with $n=1$, $\lambda_1=\lambda_2=\lambda_3=1$, and 
several values of $b_{\infty}$. {\it Right panel:} The 
relative difference $\Delta_q=(q_G-q_W)/(q_G+q_W)$ of the individual contributions to the
topological 
charge $q=q_G+q_W=1$ is shown $vs.$ 
the coupling constant 
$\lambda_0$ for the same solutions.
}}
\label{fig4}
\end{center}
\end{figure}  
%
which, as expected, is a $curl$.  
In principle, 
the value of the integral of \re{RSTa}
could be different from the integral of $\vr_0$, since it encodes a contribution of the gauge potential.
However, in contrast with the $2+1$ dimensional system endowed with Chern-Simons dynamics\footnote{
In $2+1$ dimensions, $a_\infty$ was related to $b_\infty$, which provided a lever to pass from one solution to another, resulting in the evolution of $q$, while
  $a_\infty=-n$
in $3+1$ dimensions.
}, studied in the
previous Section, 
this contribution vanishes for
regular configurations, and
the integral of \re{RSTa}
is still equal to the baryon number,
$q=n$,
 for $any$ value of the varying the coupling constants, 
in particular for any  $\la_0$.   
At the same time,  
the contribution
of the 
gauge invariant terms $q_G$ and $q_W$
 to the topological charge 
is model dependent, 
as confirmed by our numerical results\footnote{
Different from the previous work \cite{Piette:1997ny,Radu:2005jp},
the sigma-model constraint
$R^2+S^2+T^2=1$
is imposed here by using the Lagrange multiplier method, as explained $e.g.$ in \cite{Rajaraman,Radu:2008pp}.
The solutions with 
$b_{\infty} \neq 0$
are rotating \cite{Radu:2005jp}, possessing an angular momentum
$J=4\pi \lambda_0 Q_e$.
The mass/energy 
$E$ 
and
the angular momentum
$J$
are computed from the components 
$T_{00}$
and 
$T_{i0}$,
respectively,
of the energy momentum tensor
(which includes also the contribution of the Maxwell field).}.

The dependence of the mass/energy $E$ and of the electric charge $Q_e$
of the spinning Skyrmions on $\la_0$ is displayed in Figure \ref{fig4} (left panel). As expected, for fixed coupling constants $\lambda_i$ and winding number $n$,
$E$ increases with increasing electrostatic potential $b_{\infty}$ for all values of $\la_0$. 
In the gauge decoupling limit $\la_0\to\infty$, 
these are the uncharged, spinning Skyrmions characterised by $b_{\infty}$, found in Ref~\cite{Battye:2005nx}.

The difference
between the  `$bare$' charge  $q_G$
and the `$compensating$' one 
 $ q_W$ 
is displayed in Figure \ref{fig4} (right panel) as a function of $\lambda_0$. 
We see that the influence of the gauge field is to cause this quantity 
to have positive, zero and negative values in distinct theories.
For large  $\lambda_0$, the contribution of  $q_W$ is negligible, with $q_G \to q$.
However, $q_G$ decreases with $\lambda_0$, with $q_W$ dominating for small $\lambda_0$.
We note that for very small values of $\la_0$, when the difference between $q_G$ and $q_W$ is vanishing 
(or even becomes negative, as seen from Figure \ref{fig4} (right panel)), 
the energy of the Skyrmion (as seen in Figure \ref{fig4} (left panel)), is finite.

\section{Summary}
We have studied two $SO(2)$ gauged Skyrme systems, the $O(3)$ and $O(4)$ sigma models in $2+1$ and $3+1$ dimensions, respectively. In the first case, the system can be endowed with
Chern-Simons dynamics, while in the second case, not. The main message here is to highlight the contrast when Chern-Simons dynamics is present or absent.

When Chern-Simons dynamics is present, two related phenomena are observed. {\bf i)} The mass of the soliton can both increase and decrease with increasing global charge (electric or spin),
and {\bf ii)} The topological charge can evolve through positive to zero to negative values, for various solutions characterised by $a_\infty$, the asymptotic value of the magnetic field
defining the topological charge. This evolution is contingent on the dependence of $b_\infty$, the asymptotic value of the electric field, this relation being dependent on the
intertwining of the electric and magnetic fields in the definition of the Chern-Simons density. This is clearly demonstrated in the $2+1$ dimensional system studied here,
for the $SO(2)$ gauged Skyrmion, in a model with both Maxwell and Chern-Simons dynamics, and both with and without a quartic kinetic Skyrme term.

Interestingly, the energetically favoured configurations appear to occur for smaller values of the topological charge, with progressively stronger coupling $\tau$ of the Skyrme term.
Indeed, when $\tau=0$ the minimum energy configurations coincide with the winding number $n$ (see Figure~{\bf 3}), while for very large values of $\tau$ the energetically favoured configuration tends to the
$n/2$ (see quantitative details in \cite{Francisco}.)       

In the $3+1$ dimensional system studied, where there is no Chern-Simons dynamics, these phenomena are absent. 
The mass of the soliton increases monotonically with increasing
the electric charge charge, and the topological charge is fixed. 
Moreover, changing the theory by varying a coupling constant does not lead 
to changing values of the topological charge. We suggest that the phenomena of non-standard mass-electric charge/spin~\footnote{This phenomenon is observed also in (gauged) Higgs theories, where the Chern-Simons (CS)
dynamics is supplied by Higgs--CS~\cite{Tchrakian:2015pka} (HCS) terms, reported in Refs.~\cite{Navarro-Lerida:2013pua,Navarro-Lerida:2014rwa}.} and evolving topological charge
observed in the $2+1$ dimensional example are absent in the $3+1$ dimensional model here, because of the absence of Chern-Simons dynamics in the latter case.

The challenge is to supply an example of a gauged Skyrme model in $3+1$ dimensions, 
where a Chern-Simons term is defined. Such examples are considered in Ref~\cite{Tchrakian:2015pka}. In
the case of $SO(2)$ gauging considered in this work, such a Skyrme-CS term was considered, 
but it turned out that that term vanished under symmetry imposition. The challenge thus is to
consider gauging with a higher (appropriate) gauge group, but this is a subject of another investigation.

\medskip
\medskip
\noindent {\large\bf Acknowledgements}
\\
We are very grateful to V.~A. Rubakov for valuable and penetrating comments. 
E.R. gratefully acknowledges the support of DIAS. 
The work of E.R. was also partially supported by the Portuguese national
funding agency for science, research and technology
(FCT), within the Center for Research and Development
in Mathematics and Applications (CIDMA), project UID/MAT/04106/2019,
and  
also  by the  H2020-MSCA-RISE-2015 Grant No.  StronGrHEP-690904, 
the H2020-MSCA-RISE-2017 Grant No. FunFiCO-777740,
and by the CIDMA project UID/MAT/04106/2013. 
The numerical computations in Section 3 were
performed at the BLAFIS cluster in Aveiro University.

\begin{small}

\end{small}

\end{document}